# Scalar Spin of Elementary Fermions

A. Jourjine

FG CTP
Hoffmann Str. 6-8
01281 Dresden
Germany

**Abstract**

We show that, using the experimentally observed values of CKM and PMNS mixing matrices, all known elementary fermions can be assigned a new quantum number, the scalar spin, in a unique way. This is achieved without introduction of new degrees of freedom. The assignment implies that tau-neutrino should be an anti-Dirac spinor, while mu-tau leptons and charm-top, strange-bottom quarks form Dirac-anti-Dirac scalar spin doublets. The electron and its neutrino remain as originally described by Dirac.

## 1. Introduction

The origin of the quark and lepton mixing matrices and the difference in their textures has been a long-standing puzzle in elementary particle physics. Its resolution remains elusive and it is one of the top three mysteries of the Standard Model [1, 2, 3, 4]. Explanations of the textures of mixing matrices usually employ introduction of extra dynamics using either extra gauge or discrete degrees of freedom or additional space-time dimensions. Most common approaches use extra degrees of freedom with a symmetry that is broken down to some discrete subgroup. By exhaustive search of all discrete symmetry groups it is possible to find a reasonably good fit with the experimentally observed values of the lepto-quark mixing matrices [5]. However, the many parameters that are brought along make such fits less satisfactory. A number of attempts have been also made to derive a unified framework for lepto-quark mixing. Some recent work explores quark-lepton complementarity [6], TBM-Cabibbo mixing [7], the use of $SU(5)$ GUT vector fermions [8], and of "yukawaons" in a version of the "flavon" approach [9].

In this letter we derive a common representation for the quark CKM and the lepton PMNS mixing matrices without introduction of additional degrees of freedom. The difference in textures of the two matrices appears as a result of assignment of lepton and quark pairs to different multiplets of a two element discrete symmetry. The symmetry is not present in the Standard Model (SM). It appears if, instead of the Dirac spinors, we use a bi-spinor[1] representation of fermions discovered by Ivanenko and Landau [10] and further developed in [11, 12].

When Dirac degrees of freedom are extracted from bi-spinors the symmetry, called scalar spin, appears automatically as the remnant of the second Lorentz transformation invariance of bi-spinors. However, it acts in the generation space. We show that within the context of bi-spinor gauge theory [13, 14], mixing matrix textures for both quarks and leptons are essentially unique. Alternatively, the results can be viewed as derivation from known lepto-quark mixing matrix textures of unique scalar spin multiplet assignments to all elementary fermions. This paper expands the results in [13, 14, 15] about the common form of mixing for leptons and quarks.

---

[1] In the literature bi-spinors are also referred to as Ivanenko-Landau-Kähler spinors or Dirac-Kähler spinors.

## 2. Scalar Spin and Lepto-Quark Flavor Mix

For convenience, we will work with both three and four generations of elementary fermions, assuming massive Dirac neutrinos. The interplay between three and generations will be made clear below.

After spontaneous symmetry breakdown the free-field lepto-quark part of the SM or SM4 Lagrangian with massive Dirac neutrinos is given by

$$\mathcal{L}_{SM} = \mathcal{L}_q + \mathcal{L}_l, \tag{1}$$

$$\mathcal{L}_q = \overline{Q}_i^A (i\partial) Q_i^A + \overline{u}_R^A (i\partial) u_R^A + \overline{d}_R^A (i\partial) d_R^A - \left( \overline{Q}_1^A \mathrm{M}_u^{AB} u_R^B + \overline{Q}_2^A \mathrm{M}_d^{AB} d_R^B + c.c. \right), \tag{2}$$

$$\mathcal{L}_l = \overline{E}_i^A (i\partial) E_i^A + \overline{\nu}_R^A (i\partial) \nu_R^A + \overline{e}_R^A (i\partial) e_R^A - \left( \overline{E}_1^A \mathrm{M}_\nu^{AB} \nu_R^B + \overline{E}_2^A \mathrm{M}_e^{AB} e_R^B + c.c. \right), \tag{3}$$

where $Q_i^A = \{u_L^A, d_L^A\}$, $E_i^A = \{\nu_L^A, e_L^A\}$, $A = 1,2,3$ or $A = 1,\cdots,4$ denotes multiple generations of left-handed $SU(2)_L$ doublets for quarks and leptons, while generations of right-handed quarks and leptons ($u_R^A$, $d_R^A$), ($\nu_R^A$, $e_R^A$) are $SU(2)_L$ singlets. We suppress the $SU(3)_C$ dependencies, since they play no role in the following.

In the SM as well as in the SM4 the mass matrices $\mathrm{M}_{u,d,\nu,e}^{AB}$ are arbitrary complex matrices. $\mathrm{M}_{u,d,\nu,e}^{AB}$ give rise to mass spectrum and to flavor mixing. Masses are given by the eigenvalues of $\mathrm{M}_{u,d,\nu,e}^{AB}$. Flavor mixing is described by two mixing matrices: $V_{CKM} = U_L D_L^+$ in the quark sector and $U_{PMNS} = E_L N_L^+$ in the lepton sector, where $(U_L, D_L)$ are transformations for up/down quarks that transform the $u_L^A$, $d_L^A$ fields in (2) into mass bases $u_{mL}^A = (U_L)^{AB} u^B{}_L$, $d_{mL}^A = (D_L)^{AB} d^B{}_L$ so that (2) becomes flavor-diagonal. Lepton mixing matrix is defined analogously with the help of $E_L$, $N_L$ that define mass basis transforms that make (3) flavor diagonal: $e_{mL}^A = (E_L)^{AB} e^B{}_L$, $\nu_{mL}^A = (N_L)^{AB} \nu^B{}_L$.

Since $\mathrm{M}_{u,d,\nu,e}^{AB}$ are arbitrary, both mass spectrum and the mixing parameters in $V_{CKM}$, $U_{PMNS}$ of the SM are arbitrary. One of the most enduring puzzles of the SM is that both its spectrum and mixing seem to exhibit patterns. Within a single generation of lepto-quarks there is a clear exponential dependence of mass on the "size" of the gauge group of symmetry with quarks being the heaviest and neutrinos the lightest. Also the mass splitting between the members of $SU(2)_L$ doublets generally depends on the size of the group. At the same time there is a pronounced difference in texture of mixing matrices for quarks and leptons: while for quarks mixing of the first two generations dominates $V_{CKM}$, for leptons, after the recent discovery of large $\sin\theta_{13}$, values of $U_{PMNS}$ are of the same order of magnitude.

In this paper we explore a possible solution to the mixing puzzle within the context of bi-spinor gauge theory, where fermionic degrees of freedom are described by bi-spinors instead of the standard Dirac spinors [13]. Bi-spinor gauge theory has a number of interesting features not present in the SM or any of its extensions. For example it allows explicit mass terms for fermions in bi-fundamental representations and a realization of supersymmetry that places the observed fermions and bosons in supersymmetry multiplets, thus possibly explaining non-observation of supersymmetric partners of the SM particles. The current status of quantum field theory of bi-spinors and its perturbation theory is described in [16]. Leaving calculation of mass hierarchy of lepto-quarks till a follow-up publication, here we will concentrate on the mixing puzzle.

In bi-spinor gauge theory the free-field Lagrangian for lepto-quarks comes in three possible forms, involving either a single generation or a generation pair. It is given by

$$\mathcal{L}_{b-SM} = \widetilde{\mathcal{L}}_q + \widetilde{\mathcal{L}}_l, \tag{4}$$

$$\widetilde{\mathcal{L}}_q = \overline{\overline{Q}}_i^A (i\partial) Q_i^A + \overline{\overline{u}}_R^A (i\partial) u_R^A + \overline{\overline{d}}_R^A (i\partial) d_R^A - \left( \overline{\overline{Q}}_1^A \mathrm{M}_u^{AB} u_R^B + \overline{\overline{Q}}_2^A \mathrm{M}_d^{AB} d_R^B + c.c. \right), \tag{5}$$

$$\widetilde{\mathcal{L}}_l = \overline{\overline{E}}_i^A (i\partial) E_i^A + \overline{\overline{v}}_R^A (i\partial) v_R^A + \overline{\overline{e}}_R^A (i\partial) e_R^A - \left( \overline{\overline{E}}_1^A \mathrm{M}_\nu^{AB} v_R^B + \overline{\overline{E}}_2^A \mathrm{M}_e^{AB} e_R^B + c.c. \right), \tag{6}$$

where $\overline{\overline{Q}}^A = \Gamma^{AB} \overline{Q}^B$, and $\Gamma^{AB} = diag(1, 0)$ for Dirac spinors, $\Gamma^{AB} = diag(0, -1)$ for anti-Dirac spinors, and $\Gamma^{AB} \equiv \sigma_3^{AB} = diag(1, -1)$ for a Dirac-anti-Dirac doublet spinors. Dirac and anti-Dirac spinors describe a single generation each, while Dirac-anti-Dirac doublet describes two generations, but a single elementary fermion. Any other free-field fermionic Lagrangian can be formed by a adding together arbitrary number of generations the three types above.

Like in the SM, the interacting bi-spinor gauge theory is obtained by minimally gauging the free-field theory. Thus, the only formal difference between Lagrangian (1-3) and Lagrangian (4-6) and between the corresponding interacting theories is that some generations in bi-spinor theory contribute to the Lagrangians with the negative sign. We will now show that this modification is sufficient to explain the difference in the textures of mixing of quarks and leptons in a unique way.

We will begin with the four-generation case, eventually reducing the number of generations to three. The key observation is that in bi-spinor gauge theory explicit mass matrices $\mathrm{M}_{u,d}$, $\mathrm{M}_{l,\nu}$ are not arbitrary but have a specific form [16]. Generically,

$$\mathrm{M} = m\, B_1\, \mathcal{M}\, B_2. \tag{7}$$

where $m$ is a parameter with dimension of mass, $B_a \in U(2) \oplus U(2)$, $a = 1, 2$ are $4 \times 4$ block-diagonal matrices with the upper-left blocks of which mix only $A = 1, 2$, while the lower-right blocks mix only $A = 3, 4$. Factors $B_a$ are arbitrary. They have the same form for both quark and lepton sectors

$$B_a = \begin{pmatrix} U_1^a & 0 \\ 0 & U_2^a \end{pmatrix}, \qquad U_k^a = \begin{pmatrix} x_k^a & y_k^a \\ z_k^a & w_k^a \end{pmatrix} \in U(2), \qquad a,k = 1,2. \tag{8}$$

Dimensionless matrix $\mathcal{M}$ is also a direct sum of two two-dimensional matrices but now the first summand mixes generations 1 and 3 only, while the second summand mixes generations 2 and 4

$$m\,\mathcal{M} = m_1\, \mathcal{M}_R^{(p)} \oplus m_2\, \mathcal{M}_R^{(q)}, \qquad p,q = 1,2, \tag{9}$$

where $\mathcal{M}_R^{(1)}$ is diagonal and $\mathcal{M}_R^{(2)} \in U(1,1)$

$$\mathcal{M}_R^{(2)} = \begin{pmatrix} c_\lambda & s_\lambda \\ s_\lambda & c_\lambda \end{pmatrix}, \qquad s_\lambda = \sinh\lambda, \quad c_\lambda = \cosh\lambda, \quad c_\lambda^2 - s_\lambda^2 = 1. \tag{10}$$

Requiring that fermions have physical masses results in that there are only four possible cases for $\mathcal{M}$ given by

$$m\mathcal{M} = m_1 \mathcal{M}_R^{(1)} \oplus m_2 \mathcal{M}_R^{(1)\prime}, \tag{11}$$

$$m\mathcal{M} = m_1 \mathcal{M}_R^{(1)} \oplus m_2 \mathcal{M}_R^{(2)}, \tag{12}$$

$$m\mathcal{M} = m_2 \mathcal{M}_R^{(2)} \oplus m_1 \mathcal{M}_R^{(1)}, \tag{13}$$

$$m\mathcal{M} = m_1 \mathcal{M}_R^{(2)} \oplus m_2 \mathcal{M}_R^{(2)\prime}, \tag{14}$$

where prime denotes a matrix with different non-zero entries. The possible mass matrices are listed in order of increasing mass degeneracy. The first case has 4 independent mass parameters, the second and the third three mass parameters, while the fourth has two mass parameters. In the limiting case $m_1 = m_2$ in (14) $\mathcal{M} \subset U(2,2)$.

Mass degeneracy induced by $\mathcal{M}_R^{(2)}$ reflects the fact that this case describes a single 8-component Dirac-anti-Dirac particle, consisting of generation doublet of two algebraic Dirac spinors labeled by an additional quantum number, called scalar spin [16]. The degeneracy is lifted at one-loop level by interactions. The computation of the lifting at one loop will be presented elsewhere.

Having listed all possible mass terms, we turn to the diagonalization procedure. In the SM the diagonalization procedure is a linear transformation in the field generation space. Diagonalization is always possible, since after representing mass matric in its polar decomposition form one can always redefine away the unitary factors, which is possible because the free field kinetic term bilinear form corresponds to the unit matrix that commutes with the unitary matrices used in the field redefinitions.

Mixing matrices in bi-spinor gauge theory are defined exactly the same as in the SM. The quark mixing matrix is defined as

$$V = U_L D_L^+, \tag{15}$$

where $U_L, D_L$ are mass basis transforms for the fields in (5)

$$u_{mL}^A = (U_L)^{AB} u^B{}_L, \qquad d_{mL}^A = (D_L)^{AB} d^B{}_L. \tag{16}$$

The lepton mixing matrix is defined analogously as

$$U = E_L N_L^+, \tag{17}$$

where $E_L, N_L$ define lepton mass basis transforms for the fields in (6)

$$e_{mL}^A = (E_L)^{AB} e_L^B, \qquad v_{mL}^A = (N_L)^{AB} v_L^B. \tag{18}$$

In the SM transformation to mass basis is defined as a transformation that decouples the fields of different generations. After the transformation free field fermionic SM Lagrangian becomes a sum of four Lagrangians each containing the fields for one of the four generations. In bi-spinor theory the situation is somewhat different. While for diagonal mass matrix summand $\mathcal{M}_R^{(1)}$ in (9) the diagonalization means exactly the same as in SM, for the $\mathcal{M}_R^{(2)}$ summand in (9) such diagonalization is impossible, because the corresponding kinetic term bilinear matrix $\sigma_3 = diag(1,-1)$ (it would be the unit matrix for the SM with two generations) does not commute with the transformations that diagonalize mass matrices.

Therefore, the definition of diagonalization in the present case has to be modified. Instead of insisting on separating the fermionic free field Lagrangian into four independent terms, for $\mathcal{M}_R^{(2)}$ case we will define diagonalization as the transformation that diagonalizes the equations of motion. This definition is sufficient for definition of mass eigenstates [16].

The transformation that decouples equations of motion for $\mathcal{M}_R^{(2)}$ case is given by

$$W^{(2)} = \frac{1}{\sqrt{2}} \begin{pmatrix} 1 & -1 \\ 1 & 1 \end{pmatrix}, \tag{19}$$

where $W$ mixes either indexes 1 and 3 or 2 and 4. Note that the order in which stripping of the unitary factors and application of (16, 18) is applied is fixed. First comes stripping of the unitary factors in (7) and then transformation (19). For $\mathcal{M}_R^{(1)}$ case we may use the unit matrix

$$W^{(1)} = \begin{pmatrix} 1 & 0 \\ 0 & 1 \end{pmatrix}. \tag{20}$$

Therefore, we obtain the complete generic diagonalizing transformation that corresponds to four possible mass matrices (11-14) is given by

$$T^{(p,q)(r,s)} = U_L D_L^+ = \left(W_U^{(p,q)} \tilde{U}_L\right)\left(W_D^{(r,s)} \tilde{D}_L\right)^+, \quad p,q,r,s = 1,2,$$

where $\tilde{U}_L, \tilde{D}_L$ are block-diagonal

$$\tilde{U}_L = \begin{pmatrix} U_1 & 0 \\ 0 & U_2 \end{pmatrix} \in U(2) \oplus U(2),$$

$$\tilde{D}_L = \begin{pmatrix} D_1 & 0 \\ 0 & D_2 \end{pmatrix} \in U(2) \oplus U(2), \quad U_k, D_k \in U(2). \tag{21}$$

A convenient expression for $W^{(p,q)}$ is given if we swap generation 2 and 3. Then instead of mixing generations 1 and 3 or 2 and 4 matrices $\mathcal{M}_R^{(k)}$ mix generations 1 and 2 or 3 and 4. After such generation swap $W^{(p,q)}$ becomes block-diagonal

$$W^{(p,q)} = \begin{pmatrix} \mathcal{M}_R^{(p)} & 0 \\ 0 & \mathcal{M}_R^{(q)} \end{pmatrix}. \tag{22}$$

Since $W \equiv \tilde{U}_L \tilde{D}_L^+ \in U(2) \oplus U(2)$ is arbitrary we can write down $T_Q$, the generic quark, or $T_L$, the generic lepton mixing matrix as

$$T_{Q,L}^{(p,q)(r,s)} = W^{(p,q)} W_{Q,L} \left(W^{(r,s)}\right)^+, \quad p,q,r,s = 1,2, \quad W_{Q,L} = \begin{pmatrix} U_1^{Q,L} & 0 \\ 0 & U_2^{Q,L} \end{pmatrix}, \tag{23}$$

where the upper-left block of $W_{Q,L}$ mixes generations 1 and 2, while its lower-right block mixes generations 3 and 4. Matrix $W^{(p,q)}$ also satisfies $W^{(p,q)} \in U(2) \oplus U(2)$. However, in (23) the first block in $W^{(p,q)}$ mixes generations 1 and 3, while the second generations 2 and 4. Explicitly, the four possible matrices $W^{(p,q)}$ are given by

$$W^{(0,0)} = \begin{pmatrix} 1 & 0 & 0 & 0 \\ 0 & 1 & 0 & 0 \\ 0 & 0 & 1 & 0 \\ 0 & 0 & 0 & 1 \end{pmatrix}, \qquad W^{(0,1)} = \begin{pmatrix} 1/\sqrt{2} & 0 & -1/\sqrt{2} & 0 \\ 0 & 1 & 0 & 0 \\ 1/\sqrt{2} & 0 & 1/\sqrt{2} & 0 \\ 0 & 0 & 0 & 1 \end{pmatrix},$$

(24)

$$W^{(1,0)} = \begin{pmatrix} 1 & 0 & 0 & 0 \\ 0 & 1/\sqrt{2} & 0 & -1/\sqrt{2} \\ 0 & 0 & 1 & 0 \\ 0 & 1/\sqrt{2} & 0 & 1/\sqrt{2} \end{pmatrix}, \qquad W^{(1,1)} = \begin{pmatrix} 1/\sqrt{2} & 0 & -1/\sqrt{2} & 0 \\ 0 & 1/\sqrt{2} & 0 & -1/\sqrt{2} \\ 1/\sqrt{2} & 0 & 1/\sqrt{2} & 0 \\ 0 & 1/\sqrt{2} & 0 & 1/\sqrt{2} \end{pmatrix}.$$

It follows from (23) that altogether there are 16 possible types of mixing matrices in 4-generation bi-spinor theory that differ in texture. They are all parameterized by arbitrary block-diagonal $W_{Q,L}$ For convenience, their explicit forms are listed in the Appendix.

We can now compare the $3 \times 3$ sub-matrices of the 16 matrices (23) with the experimentally observed SM $3 \times 3$ mixing textures and try to find whether any of them provide a reasonable fit. For elementary particles describable by Dirac spinors the generic form of $3 \times 3$ unitary mixing matrix has a single CP violating phase $\delta$. In the most commonly used Chau-Keung parameterization both for quarks and leptons it can be written as

$$U = \begin{pmatrix} 1 & 0 & 0 \\ 0 & c_{23} & s_{23} \\ 0 & -s_{23} & c_{23} \end{pmatrix} \begin{pmatrix} c_{13} & 0 & s_{13}e^{-i\delta} \\ 0 & 1 & 0 \\ -s_{13}e^{i\delta} & 0 & c_{13} \end{pmatrix} \begin{pmatrix} c_{23} & s_{23} & 0 \\ -s_{23} & c_{23} & 0 \\ 0 & 0 & 1 \end{pmatrix}$$

(25)

$$= \begin{pmatrix} c_{12}c_{13} & s_{12}c_{13} & s_{13}e^{-i\delta} \\ -s_{12}c_{23} - c_{12}s_{23}s_{13}e^{i\delta} & c_{12}c_{23} - s_{12}s_{23}s_{13}e^{i\delta} & s_{23}c_{13} \\ s_{12}s_{23} - c_{12}c_{23}s_{13}e^{i\delta} & -c_{12}s_{23} - s_{12}c_{23}s_{13}e^{i\delta} & c_{23}c_{13} \end{pmatrix}.$$

The entries of quark CKM and lepton PMNS matrices in the SM are denoted as

$$V_{CKM} = \begin{pmatrix} V_{ud} & V_{us} & V_{ub} \\ V_{cd} & V_{cs} & V_{cb} \\ V_{td} & V_{ts} & V_{tb} \end{pmatrix}, \qquad U_{PMNS} = \begin{pmatrix} U_{e1} & U_{e2} & U_{e3} \\ U_{\mu 1} & U_{\mu 2} & U_{\mu 3} \\ U_{\tau 1} & U_{\tau 2} & U_{\tau 3} \end{pmatrix}. \qquad (26)$$

Determined from the four experimentally measured Volfenstein parameters $\lambda$, $A$, $(\bar{\rho}+i\bar{\eta})$,

$$\lambda = 0.22457^{+0.00186}_{-0.00014}, \quad A = 0.823^{+0.012}_{-0.033}, \quad \bar{\rho} = 0.1289^{+0.0176}_{-0.0094}, \quad \bar{\eta} = 0.348^{+0.012}_{-0.012}, \quad (27)$$

the three mixing angles and the CP violating Kobayashi-Maskawa phase $\delta_Q$ for quarks, defined as [17]

$$\sin\theta_{12} = \lambda = \frac{|V_{us}|}{\sqrt{|V_{us}|^2 + |V_{us}|^2}}, \quad \sin\theta_{23} = A\lambda^2 = \lambda\frac{|V_{cb}|}{|V_{us}|},$$

$$\sin\theta_{13}e^{i\delta_Q} = V_{ub}^* = \frac{A\lambda^3(\bar{\rho}+i\bar{\eta})\sqrt{1-A^2\lambda^4}}{\sqrt{1-\lambda^2}\left[1-A^2\lambda^4(\bar{\rho}+i\bar{\eta})\right]},$$

(28)

are given by global fit analysis [18] as

$$\sin\theta_{12} = 0.22457^{+0.00186}_{-0.00014}, \quad \sin\theta_{23} = 0.0415^{+0.00060}_{-0.0016}, \quad \sin\theta_{13} = 0.00355^{+0.00016}_{+0.00013},$$

$$\delta_Q = 69.7^{\circ\,+1.96^\circ}_{\,-3.26^\circ}.$$

(29)

For leptons the most recently measured values of the mixing angles are [19, 20]

$$\sin^2\theta_{12} = 0.312^{+0.018}_{-0.015}, \quad \sin^2\theta_{23} = 0.42^{+0.08}_{-0.03}, \quad \sin^2\theta_{13} = 0.0251 \pm 0.0034, \qquad (30)$$

while their recent best fit [21] is (we quote for normal hierarchy)

$$\sin^2\theta_{12} = 0.307^{+0.018}_{-0.016}, \quad \sin^2\theta_{23} = 0.386^{+0.024}_{-0.021}, \quad \sin^2\theta_{13} = 0.0241^{+0.0025}_{-0.0025},$$

$$\delta_L = \pi(1.08^{+0.28}_{-0.31}) = 194^{\circ\,+50^\circ}_{\,-56^\circ}.$$

(31)

The leptonic phase $\delta_L$ presently cannot be measured experimentally, however, global analysis indicates a $1\sigma$ preference for $\delta_{PMNS} = \pi$ and arbitrary value at $2\sigma$ [21].

To show how much more precise the determination of the CKM matrix under the assumption of $3\times 3$ SM unitarity is, we write down the most recent global fit for absolute values of CKM matrix [18]

$$|V_{CKM}| = \begin{pmatrix} 0.974452 {}^{+0.000033}_{-0.000432} & 0.22457 {}^{+0.00186}_{-0.00014} & 0.00355 {}^{+0.00016}_{-0.00013} \\ 0.22443 {}^{+0.00186}_{-0.00015} & 0.973607 {}^{+0.000069}_{-0.000445} & 0.04151 {}^{+0.00056}_{-0.00115} \\ 0.00875 {}^{+0.00016}_{-0.00031} & 0.04073 {}^{+0.00055}_{-0.00113} & 0.999132 {}^{+0.000047}_{-0.000024} \end{pmatrix}, \qquad (32)$$

and its directly measured values from [17]

$$|V_{CKM}| = \begin{pmatrix} 0.9745 \pm 0.00022 & 0.2252 \pm 0.0009 & 0.00415 \pm 0.00049 \\ 0.230 \pm 0.011 & 1.006 \pm 0.023 & 0.0409 \pm 0.0011 \\ (X) & (X) & 0.89 \pm 0.07 \end{pmatrix}, \qquad (33)$$

where $V_{td}$, $V_{ts}$ in (33) are not directly measurable and hence marked by (X). They can be calculated via box diagrams assuming $3\times 3$ unitarity and $V_{tb} = 1$ to produce $V_{td} = 0.0084 \pm 0.0006$, $V_{ts} = 0.0429 \pm 0.0026$. For leptons we have from direct measurements of absolute values

$$|U_{PMNS}| = \begin{pmatrix} 0.819 {}^{+0.010}_{-0.012} & 0.552 {}^{+0.017}_{-0.014} & 0.158 {}^{+0.010}_{-0.011} \\ 0.440 {}^{+0.035}_{-0.016} & 0.630 {}^{+0.039}_{-0.016} & 0.640 {}^{+0.061}_{-0.023} \\ 0.368 {}^{+0.041}_{-0.018} & 0.547 {}^{+0.045}_{-0.018} & 0.752 {}^{+0.052}_{-0.019} \end{pmatrix}, \qquad (34)$$

The $|U_{PMNS}|$ is close to absolute values of the tri-maximal matrix $U_{TBM}$

$$U_{TBM} = \begin{pmatrix} \sqrt{2/3} & \sqrt{1/3} & 0 \\ \frac{1}{\sqrt{2}} \cdot (-\sqrt{1/3}) & \frac{1}{\sqrt{2}} \cdot \sqrt{2/3} & \frac{1}{\sqrt{2}} \cdot 1 \\ \frac{1}{\sqrt{2}} \cdot \sqrt{1/3} & -\frac{1}{\sqrt{2}} \cdot \sqrt{2/3} & \frac{1}{\sqrt{2}} \cdot 1 \end{pmatrix} = \begin{pmatrix} 0.82 & 0.58 & 0 \\ -0.41 & 0.58 & 0.71 \\ 0.41 & -0.58 & 0.71 \end{pmatrix}, \qquad (35)$$

where to make visible the $2\times 2$ unitary matrix in the upper-left corner of $U_{TBM}$ we extracted factor $1/\sqrt{2}$ from its values.

We can now examine the 16 possible $4\times 4$ unitary matrices (23) that are listed in the Appendix. We seek two $4\times 4$ matrices such that after cutoff of one of the generations the resulting (non-unitary) $3\times 3$ matrices approximate the experimental data in the best way. It is not to difficult to find that there is an unequivocal best fit choice for both quarks and leptons given by

$$\widehat{V}_{CKM}^{4} \equiv (1,1)(1,1) = \frac{1}{2}\begin{pmatrix} x_1+x_2 & y_1+y_2 & x_1-x_2 & y_1-y_2 \\ z_1+z_2 & w_1+w_2 & z_1-z_2 & w_1-w_2 \\ x_1-x_2 & y_1-y_2 & x_1+x_2 & y_1+y_2 \\ z_1-z_2 & w_1-w_2 & z_1+z_2 & w_1+w_2 \end{pmatrix}, \quad (36)$$

$$\widehat{U}_{PMNS}^{4} \equiv (1,0)(0,0) = \begin{pmatrix} \hat{x}_1 & \hat{y}_1 & 0 & 0 \\ \hat{z}_1/\sqrt{2} & \hat{w}_1/\sqrt{2} & -\hat{z}_2/\sqrt{2} & -\hat{w}_2/\sqrt{2} \\ 0 & 0 & \hat{x}_2 & \hat{y}_2 \\ -\hat{z}_1/\sqrt{2} & -\hat{w}_1/\sqrt{2} & -\hat{z}_2/\sqrt{2} & -\hat{w}_2/\sqrt{2} \end{pmatrix}, \quad (37)$$

where the "hatted" matrix elements denote a different choice of parameters of the unitary block diagonal matrices in (23) and where we multiplied the last row of $\widehat{U}_{PMNS}^{4}$ by $-1$. Each block in (23) is a $U(2)$ matrix with one real parameter and three phases, which can be represented as

$$\begin{pmatrix} x & y \\ z & w \end{pmatrix} = \begin{pmatrix} \cos\theta e^{i\alpha_x} & \sin\theta e^{-i\alpha_y} \\ -\sin\theta e^{-i\alpha_z} & \cos\theta e^{i\alpha_w} \end{pmatrix}, \alpha_x + \alpha_y + \alpha_z + \alpha_w = 0. \quad (38)$$

From (36, 37) we obtain two $3\times 3$ mixing matrices $V_{b-CKM}$ and $U_{b-PMNS}$ by elimination of the $A=3$ rows and columns via spinbein cutoff. The reason why specifically the third rows and columns are eliminated is simple. It provides the best fit between $V_{b-CKM}$, $U_{b-PMNS}$ and experimental data. The elimination is in fact done in a generally covariant way by imposing a generally covariant constraint of the type $\det\Psi = 0$, where $\Psi$ is a quark or lepton bi-spinor field [16, 22, 23]. We arrive at the final form of $3\times 3$ mixing matrices for the bi-spinor gauge theory with Dirac neutrinos

$$V_{b-CKM} = \frac{1}{2}\begin{pmatrix} x_1+x_2 & y_1+y_2 & y_1-y_2 \\ z_1+z_2 & w_1+w_2 & w_1-w_2 \\ z_1-z_2 & w_1-w_2 & w_1+w_2 \end{pmatrix}, \quad (39)$$

$$U_{b-PMNS} = \begin{pmatrix} \hat{x}_1 & \hat{y}_1 & 0 \\ \hat{z}_1/\sqrt{2} & \hat{w}_1/\sqrt{2} & -\hat{w}_2/\sqrt{2} \\ -\hat{z}_1/\sqrt{2} & -\hat{w}_1/\sqrt{2} & -\hat{w}_2/\sqrt{2} \end{pmatrix}. \quad (40)$$

Unlike the general SM mixing matrix in (25), quark bi-spinor mixing matrix has two real parameters and three phases, while lepton bi-spinor mixing matrix has two real parameters and no phases, that is, by adjustment of phases of fermion fields it can be chosen to be real. As we will see below, additional phases may appear as a result of renormalization. The parameter count can be obtained as follows.

Since we have three generations of quarks, there are six re-scaling phases available, which however may be taken to sum to zero, since the Lagrangian is invariant with regard to rescaling with the same phase. Thus, we are left with five phases available for rescaling. For quarks, since we want to preserve the form of (39) we have to fix one of the phases to ensure that after rescaling $V_{cs} = V_{tb}$, $V_{ts} = V_{cb}$ are satisfied. $V_{b-CKM}$ has seven independent entries, the phases of which depend on six phases, three from each of the two $U(2)$ that enter (23). We can define six new phases that are the phases of the last two entries in the first row plus three entries in the second row plus the first entry in the third row. The phase of $x_1 - x_2$ in (39) is then a function of the six new phases. Under rescaling of quark fields an entry $a_{ij}$ of the mixing matrix rescales to $a_{ij} e^{-i(\phi_i - \chi_j)}$. Therefore, by rescaling we can eliminate four of the seven phases. To make the parameterization the closest to (25) we will choose the phases of the upper-left $2 \times 2$ block in (39). Therefore, for quark b-CKM mixing matrix we obtain

$$V_{b-CKM} = \begin{pmatrix} \rho_x^+ & \rho_y^+ & \rho_y^- e^{i\delta_y} \\ \rho_z^+ & \rho_w^+ & \rho_w^- e^{i\delta_w} \\ \rho_z^- e^{i\delta_z} & \rho_w^- e^{i\delta_w} & \rho_w^+ \end{pmatrix}, \qquad (41)$$

where $\rho_x^+ = \left(|x_1|^2 + |x_2|^2 + |x_1||x_2|\cos(\alpha_{x_1} - \alpha_{x_2})\right)$, ..., and $\delta_y, \delta_z, \delta_w$ are three remaining independent phases, $\delta_y = \arctan(\mathrm{Im}(y_1 - y_2)/\mathrm{Re}(y_1 - y_2))$, .... Taking $\delta_w = 0$ we obtain parameterization of b-CKM that is close to that of CKM with accuracy of $O(\lambda^4) \approx 10^{-3}$, except that the sign of $V_{ts}$ for b-CKM in (39) is the opposite of $V_{ts}$ for CKM matrix in (25, 26), where $V_{cb} = -V_{ts} = A\lambda^2$. Unfortunately the difference in sign cannot be exploited at this time, since $V_{ts}$ cannot be determined directly from the experiment.

For lepton b-PMNS mixing matrix in (40) we also have four phases available for rescaling, reduced from the originally available five by requiring that the form of the matrix remains unchanged, which fixes one of is phases. At the same time it depends on four phases: the three phases of the upper-left $2 \times 2$ block plus one phase of $\hat{w}_2$. Thus b-PMNS mixing matrix can be chosen to be real by field rescaling. This implies that either $\delta_L = 0$ or $\delta_L = \pi$. It seems that global fit in (31) indicates a $1\sigma$ preference for $\delta_L = \pi$.

Having settled the issue of the number of independent parameters for b-mixing, we now can try to find the values of the entries in the $U(2)$ blocks in (23) that describe the experimental data the best. We obtain the experimentally observed textures in (25, 26) if we take for quarks and leptons

$$W_Q = \begin{pmatrix} U_1^Q & 0 \\ 0 & U_2^Q \end{pmatrix}, \qquad U_1^Q \approx U_2^Q,$$

$$W_L = \begin{pmatrix} U_1^L & 0 \\ 0 & U_2^L \end{pmatrix}, \qquad U_1^L \approx \begin{pmatrix} \sqrt{2/3} & \sqrt{1/3} \\ -\sqrt{1/3} & \sqrt{2/3} \end{pmatrix}, \qquad U_2^L \approx \begin{pmatrix} -1 & 0 \\ 0 & -1 \end{pmatrix}. \qquad (42)$$

Under the assumption (42) the resulting quark mixing matrix (39) correctly predicts that $|V_{ts}| \approx |V_{cb}|$, while leaving $|V_{td}|, |V_{ub}|$ independent. It also predicts that mixing of the third generation with the first two is suppressed. If we assume (42) with the equality sign for leptons then b-PMNS matrix reduces to the TBM matrix (35).

We now notice that the choice of (36, 37) uniquely specifies scalar spin assignments of quarks and leptons in 4-generation bi-spinor gauge theory. We conclude that $u - t'$, $c - t$, $d - b'$, and $s - b$ are scalar spin 1/2 DaD doublets. For leptons, $e$ and $e_4$ have scalar spin zero, where $e$ is a Dirac spinor, while $e_4$ is an anti-Dirac spinor. At the same time, $\mu - \tau$ form a scalar spin 1/2 Dirac-anti-Dirac doublet. Note, that DaD doublets $p-q$ in fact must be considered as manifestations of a single particle, where $p$ is a state of $p-q$ with scalar spin up, while $q$ is a state of $p-q$ with scalar spin down. (The direction in the space of scalar spin is determined by interaction with gauge fields.) Therefore, we can assign index $A = 1,\cdots,4$ to the four generations of b-gauge theory according to

$$\left(v^1, v^2, v^3, v^4\right) = \left(v_e, v_\mu, v_{e_4}, v_\tau\right), \qquad \left(e^1, e^2, e^3, e^4\right) = \left(e, \mu, e_4, \tau\right), \qquad (43)$$
$$\left(u^1, u^2, u^3, u^4\right) = \left(u, c, t', t\right), \qquad \left(d^1, d^2, d^3, d^4\right) = \left(d, s, b', b\right).$$

Note that in this assignment the conventional numbering of the fourth and the third generations are switched. However, masses of $t', b'$ and $e_4, v_{e_4}$ should not be assumed to be smaller than masses of $t, b$ and $\tau, v_\tau$, respectively. Their values are in any case irrelevant, because the dynamics of the fourth generation is cut off from the Lagrangian. The cutoff should not be confused with the well-known effect of decoupling of dynamics due to very large mass of a particle. The cut-off generation four leaves no traces in the dynamics, e.g., in loop calculations. Its presence can be detected only kinematically, through its influence on the form of mixing matrices. After the cutoff we obtain the final assignment in 3-generation bi-SM

$$\left(v_e, v_\mu, v_\tau\right) = \left(v^1, v^2, v^4\right),$$
$$\left(e, \mu, \tau\right) = \left(e^1, e^2, e^4\right),$$
$$\left(u, c, t\right) = \left(u^1, u^2, u^4\right), \qquad (44)$$
$$\left(d, s, b\right) = \left(d^1, d^2, d^4\right).$$

From (36, 37) we see that unlike the SM mixing, b-SM mixing is constrained. As a result, b-SM mixing predicts a number of relations between mixing matrix elements. If these are grossly violated, b-SM can be ruled out. Small violations could be acceptable, since in bi-spinor theory mixing matrix elements are modified after renormalization. Let us, therefore, review the experimental data on $V_{CKM}$ and $U_{PMNS}$ in detail. First, let us see how well the characteristic relations $|V_{cs}| = |V_{tb}|$, $|V_{ts}| = |V_{cb}|$ in (39) are satisfied. We should bear in mind that for comparison one must use direct measurements of mixing matrix elements or measurements where difference between $3\times 3$ unitarity of the SM and $4\times 4$ unitarity of bi-spinor mixing is expected to be insignificant. From three experiments in neutrino scattering, semileptonic, and leptonic decays $|V_{cs}| = 0.94^{+0.32}_{-0.26} \pm 0.13$, $|V_{cs}| = 0.98 \pm 0.01_{\text{exp}} \pm 0.10_{\text{theor}}$, $|V_{cs}| = 1.030 \pm 0.038$, respectively [17]. More recently CDF reports $|V_{tb}| = 0.97 \pm 0.05$, while CMS under assumption of SM unitarity obtains $|V_{tb}| = 0.98 \pm 0.04$ [24]. In the SM, as a result of $3\times 3$ unitarity, one expects $|V_{tb}|$ very close to one. Most recent global unitarity fit gives $|V_{tb}| = 0.999132^{+0.000047}_{-0.000024}$. As far as $|V_{ts}| = |V_{cb}|$ is concerned, from inclusive/exclusive semileptonic decays [25] $|V_{cb}|_{inc} = (41.96 \pm 0.45) \cdot 10^{-3}$ in the 1S subtraction scheme, while for exclusive decays from PDG [17] $|V_{cb}|_{excl} = (40.7 \pm 1.5_{exp} \pm 0.8_{theo}) \cdot 10^{-3}$. $|V_{ts}|$ presently cannot be determined directly and its determination involves the assumption of the SM unitarity. However, this assumption does not affect the value or the error significantly if unitarity violation is small, which as we will see below is indeed the case. From PDG-2012 we get $|V_{ts}| = (42.9 \pm 2.6) \cdot 10^{-3}$, obtained with partial use of SM unitarity in loop calculations and assuming $|V_{tb}| = 1$ [17]. In summary, we note that experimentally the relations $|V_{cs}| = |V_{tb}|$ and $|V_{ts}| = |V_{cb}|$ are within a standard deviation in the error bounds and at present one cannot rule out bi-spinor gauge theory on the basis of the relations.

As noted above, $V_{b-CKM}$ and $U_{b-PMNS}$ in (39, 40) are not unitary. However, they are approximately unitary and the deviation from unitarity would be difficult to detect. Presently $U_{PMNS}$ is not measured with sufficient accuracy to provide meaningful testing of the unitarity constraints. Still, from (40) we obtain

$$\det U_{b-PMNS} = \det \begin{pmatrix} \hat{x}_1 & \hat{y}_1 & 0 \\ \hat{z}_1/\sqrt{2} & \hat{w}_1/\sqrt{2} & -\hat{w}_2/\sqrt{2} \\ -\hat{z}_1/\sqrt{2} & -\hat{w}_1/\sqrt{2} & -\hat{w}_2/\sqrt{2} \end{pmatrix} = -\hat{w}_2, \qquad (45)$$

which for $\hat{w}_2 = -1$ implies $3 \times 3$ unitarity. As for $V_{b-CKM}$, from (39) we obtain the leading terms with accuracy up to $10^{-3}$

$$\det V_{CKM} = \det \begin{pmatrix} \Delta_+ x & \Delta_+ y & \Delta_- y \\ \Delta_+ z & \Delta_+ w & \Delta_- w \\ \Delta_- z & \Delta_- w & \Delta_+ w \end{pmatrix} = (\Delta_+ x \Delta_+ w - \Delta_+ y \Delta_+ z)\Delta_+ w + \Delta_+ x (\Delta_- w)^2 + \cdots, \qquad (46)$$

where $\Delta_\pm x \equiv (x_1 \pm x_2)/2, \ldots$. The uncertainty here is $\approx 10^{-1}$, which is fairly large. It is dominated by the uncertainty in the value of $V_{tb}$, since the second term contributes $\approx 10^{-3}$. Experimentally, only the first row of $V_{CKM}$ offers the best chance to make a partial test of $3 \times 3$ unitarity. Its measured value is $|V_{ud}|^2 + |V_{us}|^2 + |V_{ub}|^2 = 0.9999 \pm 0.0006$ [17]. The accuracy of this measurement is not sufficient to distinguish between the exact $3 \times 3$ unitarity SM and the approximate $3 \times 3$ unitarity of bi-spinor theory. However, increasing accuracy in determination of $V_{tb}$ by two orders of magnitude and that of $V_{cd}$ by one order of magnitude would make it possible.

Despite the encouraging hints from the experiment, there are two problems that stand in the way of treating scalar spin as physical quantity. The first problem is that we have obtained the tree level relation $U_{e3} = 0$. The second problem is that members of Dirac-anti-Dirac doublets are degenerate in mass. Both mass degeneracy and $U_{e3} = 0$ contradict observations. Let us consider the two problems in turn.

We begin with $U_{e3} = 0$. So far we provided a possible explanation of the textures for CKM and PMNS mixing matrices by assuming approximations (42). From the analysis of renormalization of the propagator and inter-generation mixing matrices in the SM [26] it follows the renormalization effects could be as high as one percent. OF course, since in the SM all Yukawa couplings are arbitrary, in the SM this result is of little significance.

The situation changes in bi-spinor SM. There the tree-level mixing matrix entries are no longer arbitrary but can be grouped according to their order of magnitude: for quarks there are four entries in CKM matrix that are on order or less of a percent in absolute value, while in PMNS matrix there is only one, the $U_{e3}$, which is approximately fifteen percent in value.

Therefore, in the bi-spinor gauge theory it is reasonable to try to explain the small observed values in mixing matrices as originating from radiative correction to tree-level values, assuming that at tree level relations (42) are exact. If radiative corrections can modify mixing of the second and third lepton generation, for example, if a $U(2)$ $\mu, \tau$ - neutrino wave-function renormalization results in

$$W^{(0,0)} = \begin{pmatrix} 1 & 0 & 0 & 0 \\ 0 & 1 & 0 & 0 \\ 0 & 0 & 1 & 0 \\ 0 & 0 & 0 & 1 \end{pmatrix} \rightarrow W_{rad}^{(0,0)} = \begin{pmatrix} 1 & 0 & 0 & 0 \\ 0 & c_\theta e^{i\alpha_1} & 0 & s_\theta e^{-i\alpha_2} \\ 0 & 0 & 1 & 0 \\ 0 & -s_\theta e^{-i\alpha_3} & 0 & c_\theta e^{i\alpha_4} \end{pmatrix}, \quad (47)$$

where $\sum \alpha_k = 0$, then 4-generation b-PMNS modifies to

$$\hat{U}_{PMNS}^4 \rightarrow \begin{pmatrix} \hat{x}_1 & \hat{y}_1 & 0 & 0 \\ \hat{z}_1/\sqrt{2} & \hat{w}_1/\sqrt{2} & -\hat{z}_2/\sqrt{2} & -\hat{w}_2/\sqrt{2} \\ 0 & 0 & \hat{x}_2 & \hat{y}_2 \\ -\hat{z}_1/\sqrt{2} & -\hat{w}_1/\sqrt{2} & -\hat{z}_2/\sqrt{2} & -\hat{w}_2/\sqrt{2} \end{pmatrix} W_{rad}^{(0,0)},$$

which results in radiatively corrected 3-generation b-PMNS

$$U_{b-PMNS} = \begin{pmatrix} \hat{x}_1 & \hat{y}_1 c_\theta e^{i\alpha_1} & \hat{y}_1 s_\theta e^{-i\alpha_2} \\ \hat{z}_1/\sqrt{2} & (\hat{w}_1 c_\theta e^{i\alpha_1} + w_2 s_\theta e^{-i\alpha_3})/\sqrt{2} & (\hat{w}_1 s_\theta e^{-i\alpha_2} - w_2 c_\theta e^{i\alpha_4})/\sqrt{2} \\ \hat{z}_1/\sqrt{2} & (\hat{w}_1 c_\theta e^{i\alpha_1} - w_2 s_\theta e^{-i\alpha_3})/\sqrt{2} & (\hat{w}_1 s_\theta e^{-i\alpha_2} + w_2 c_\theta e^{i\alpha_4})/\sqrt{2} \end{pmatrix}, \quad (48)$$

whose form is sufficient to explain the small non-zero $U_{e3}$ and all experimental results [27]. Modification of $\hat{U}_{PMNS}^4$ by multiplication of $\hat{U}_{PMNS}^4$ by $W_{rad}^{(0,0)}$ from the right is in fact a slight generalization of the proposal in [27], where one begins with the TBM PMNS matrix and then corrects it by multiplication from the right by a two parameter matrix that is a member of $1 \oplus U(2)$, which precisely the form of $W_{rad}^{(0,0)}$. In fact we obtain mixing matrix in [27] for the choice $\alpha_1 = \alpha_4 = 0$, $\alpha_2 + \alpha_3 = 0$. Of course, in absence of detailed calculations such a modification to (48) is speculative. As for the origin of the $U(2)$ blocks in (23), if instead of (42) we begin with tree-level

$$W_Q = \begin{pmatrix} U_1^Q & 0 \\ 0 & U_2^Q \end{pmatrix}, \quad U_1^Q = U_2^Q, \quad (49)$$

for quarks and

$$W_L = \begin{pmatrix} U_1^L & 0 \\ 0 & U_2^L \end{pmatrix}, \quad U_1^L = \begin{pmatrix} \sqrt{2/3} & \sqrt{1/3} \\ -\sqrt{1/3} & \sqrt{2/3} \end{pmatrix}, \quad U_2^L = \begin{pmatrix} -1 & 0 \\ 0 & -1 \end{pmatrix}, \quad (50)$$

for leptons then we notice that the form of $U_1^L, U_2^L$ is very specific. It is so specific that one suspects that there must be some deeper reason for it then a chance. Such values are typically determined by some underlying symmetries, in this case by $SU(3)_C \times SU(2)_L \times U(1)_Y$ for quarks and $SU(2)_L \times U(1)_Y$ for leptons. As a pure numerology attempt one can write down tree level $U_{1,2}^{Q,L}$ as

$$U_1^Q = \begin{pmatrix} \sqrt{1-a_Q^2} & a_Q \\ -a_Q & \sqrt{1-a_Q^2} \end{pmatrix} = U_2^Q, \quad a_Q = \left(\sqrt{f_C} + \sqrt{f_L}\right)^{-1}, \quad a_Q \approx 0.220,$$

$$U_1^L = \begin{pmatrix} \sqrt{1-a_L^2} & a_L \\ -a_L & \sqrt{1-a_L^2} \end{pmatrix}, \quad a_L = \left(\sqrt{f_L}\right)^{-1}, \quad a_L \approx 0.577 \quad (51)$$

$$U_2^L = -\begin{pmatrix} \sqrt{1-b_L^2} & b_L \\ -b_L & \sqrt{1-b_L^2} \end{pmatrix}, \quad b_L = 0,$$

where $f_C = 8$, $f_L = 3$ are the numbers of generators of semi-simple factors of the SM gauge group. Of course, the only justification for such a choice that it approximately reproduces numerically the absolute values of the CKM and PMNS matrices with the use of small integers related to gauge group of the SM: experimentally for quarks $|V_{us}| \approx 0.225$ must be compared to $a_Q \approx 0.220$ and for leptons $|U_{e2}| \approx 0.55$ must be compared to $a_L \approx 0.58$.

As far as mass degeneracy of DaD doublets is concerned, it is lifted by the interactions of the doublets with gauge fields, because the interaction term does not commute with the piece of the free Lagrangian that is proportional to $\sinh \lambda$ parameter in (10). As a result, there will be corrections to the fermion self-energy that are proportional to $\sinh \lambda$ that depend on whether scalar spin is up or down. Hence mass degeneracy is be broken by radiative corrections. Notably, the 1-loop corrections are proportional to coupling constant squared, which implies that mass splitting for quarks should be larger than that for leptons, which is indeed the case. Similar effect takes place at one loop for corrections to the $\mu - \tau$ vertex. Despite restrictions on mixing matrices, b-SM does not predict masses. Bare masses of Dirac, anti-Dirac, DaD particles and $\sinh \lambda$ are free parameters of bi-spinor gauge theory, hence absolute values of masses cannot be predicted. However, the ratio of mass difference to the average mass of a scalar spin doublet would be calculable. Such detailed calculations are beyond the scope of the paper and will be presented elsewhere.

### 3. Summary

We have shown that within the framework of bi-spinor gauge theory the measured textures of the lepto-quark mixing matrices lead to unique assignment of scalar spin multiplets to all experimentally observed elementary fermions. The result raises a number of questions, the answers to which at present are mostly lacking. The questions illuminate, however, further research that needs to be carried out in order to make scalar spin not a experimentally hypothetical but physical quantity that originates from a well defined quantum field theory. Of course, the answers to the questions can also rule out both scalar spin and bi-spinor gauge theories as alternatives to the standard gauge theories. Let us consider the most obvious questions in turn.

First we summarize. We began with free bi-spinor dynamics. It has been known for some time that, unlike in the SM, bi-spinor gauge theories with left-right asymmetry admit explicit mass terms [13]. This happens, because generically bi-spinors transform in bi-fundamental representations of the gauge group. Subsequently, it was established that the explicit mass terms are severely restricted in their form [16]. The restrictions appear when one extracts from bi-spinors the Dirac degrees of freedom via spinbein decomposition, because the free-field Lagrangian expressed in terms of algebraic Dirac spinors retains remnants of bi-spinor transformation property of bi-spinors. Thus, scalar spin appears as a residue of symmetry with regard to the Lorentz transformation applied to the second bi-spinor Dirac index.

The results presented in this paper are based on the tree-level analysis of this free-field fermionic bi-spinor Lagrangian, which differs from free-field fermionic Lagrangian of the SM by the fact that the massless part of the Lagrangian for some of generations enters the Lagrangian with the negative sign. They indicate that, using the experimental data on quark and lepton mixing, scalar spin value can be assigned to all known fermions in a unique way. Naturally, a question arises whether one can take the gauge group of the SM and construct its bi-spinor analog using minimal gauging of the bi-spinor free field Lagrangian as is done with the SM? This of course can be easily done at tree level. Assuming that full quantum field theory for bi-spinor gauge theories is constructed, what would be relation of such theory, let us call it b-SM, to the SM and what would be its phenomenological consequences? Would it be consistent with the experimental data, which the SM fits so well?

Unfortunately, at this stage in the development of quantum bi-spinor gauge theory it is not possible to answer questions, answers to which rely on loop calculations. The reason for this is that quantum field theory of bi-spinor and gauge fields differs from that of spinor and gauge fields. Therefore, a careful analysis and construction of general bi-spinor gauge theory is needed before detailed loop calculations can be carried out. Hence, here we will restrict ourselves to very general arguments. For progress on bi-spinor QFT we refer to [16].

As far as general arguments are concerned, massless or explicitly massive bi-spinor gauge theory obtained from massless or massive free-field bi-spinor theory by minimal gauging is renormalizable by power count. All of its coupling constants are dimensionless. Clearly minimal gauging would affect not only electroweak but the QCD interactions as well: the massless part of the interacting fermionic action of some generations of quarks and, separately, of leptons could enter the total action with the negative sign. Minimal gauging for such generations would add interaction terms with coupling constants negative of those in the analogous SM. Nevertheless, coupling in b-SM is as universal as in the SM: gauge fields couple to fermions with coupling constants that have the same sign relative to free-field Lagrangian.

The next natural question is about the role of Higgs field in the theory. In the SM all masses both for gauge and fermionic fields are generated as a result of the existence of non-zero vacuum expectation value of Higgs field. Bi-spinor SM seems to offer an alternative for fermions. There fermionic masses can appear as the result of normalization of spinbeins, in a process that is purely kinematical in origin, where a surrogate Higgs field doublet appears from spinbeins. How to combine the standard Higgs effect with the kinematic Higgs effect in b-SM is not clear at the moment. It is also not clear, whether the Higgs doublet should be complex or real. Supersymmetry in bi-spinor gauge theory requires that spin zero counterpart of fermionic members of a chiral supermultiplet is real [22, 23].

We emphasize, that although we formally started with four generations and in the derivation of the results the fourth generation is essential for fitting the experimental data, the fourth generation of quarks and leptons is excluded from the effective dynamics of Dirac degrees of freedom in b-SM not because of standard decoupling argument, which is based heavy mass of the fourth generation, but because the fourth generation is cut off from the dynamics by the choice of degenerate spinbein. As a result, the fourth generation does not enter the Lagrangian, except in CKM and PMNS mixing matrices. Because the fourth generation is present in the theory only kinematically and not dynamically, it does not contribute to loop integrals at all. Hence it does not have to be decoupled.

We saw above that electron and electron neutrino are the only particles in bi-spinor SM that are described by the standard Dirac spinor action. In b-SM the Dirac's quantum theory of electron would still hold. At the same time $\mu - \tau$ and their neutrinos are not described by Dirac theory but form two DaD doublets of scalar spin ½. As for quarks, all of them are members of DaD doublets, $u - \varnothing$, $c - t$, $d - \varnothing$, $s - b$, where by $\varnothing$ we symbolically denoted the fourth generation states cut off by the spinbein. Apparently, there exists a difference between the SM and bi-spinor SM. Is the difference physical? Can it be detected?

The difference is physical, because anti-Dirac (DaD) particles could couple differently to fields of integer spin than Dirac particles, but detection of the difference would not be straightforward. This is because the most pronounced differences between Dirac and anti-Dirac (DaD) particles would appear in the amplitudes where contributions of Dirac and non-Dirac particles would create an interference effect. This could be difficult to detect because of large differences in particle masses. The differences in mass would lead to suppression of the contribution to the amplitude of the lighter particle by the ration of two masses. It follows then that the interference terms would at most contribute some percentage points to the amplitudes and even less to the scattering probabilities.

But what about the precision electroweak measurements, and the S,T,U parameters [28] that are designed to detect in electroweak vacuum polarization contributions of yet unseen heavy particles? Even with the S, T, U parameters the situation presently is not so clear. This is because the hypothetical bi-spinor SM is not an extension of the SM: the propagators of anti-Dirac and DaD doublets differ from the standard Feynman propagators of Dirac particles. As a result, the S, T, U parameters for bi-spinor SM are not compatible with those of the SM. This means that to enforce the EW constraints one has to re-derive the whole machinery of oblique corrections and then compare the experimental values of new S, T, U bi-SM parameters with their theoretical predictions for bi-spinor SM. This work is in progress.

Before one carries out the construction of full b-SM and computes S, T, U parameters, however, there are two urgent problems to solve. The first problem is that we have obtained $U_{e3} = 0$. The second problem is that members of Dirac-anti-Dirac doublets are degenerate in mass. Both contradict observations. As was outlined at the end of the preceding section both $U_{e3} = 0$ and mass degeneracy should be cured by one loop corrections, the detailed calculations, however, are beyond the scope of the present work.

In summary, the results presented here can be best be considered as providing motivation for carrying out further work in rewriting the spinor part of the standard QFT in terms of bi-spinors and constructing a well-defined b-SM that includes Higgs sector. Of course, this work is conditioned on the satisfactory resolution of $U_{e3} = 0$ and DaD mass degeneracy problems.

## Appendix

The 16 possible $4\times 4$ bi-spinor mixing matrices (23) $T_Q$ for quarks or $T_L$ leptons are given by

$$T_{L,Q}{}^{(p,q)(r,s)} = W^{(p,q)} W_{L,Q} \left(W^{(r,s)}\right)^+, \qquad p,q,r,s = 1,2.$$

With the notation $T_{L,Q}{}^{(p,q)(r,s)} = (p,q)(r,s)$ their explicit forms are

$$(0,0)(0,0) = \begin{pmatrix} x_1 & y_1 & 0 & 0 \\ z_1 & w_1 & 0 & 0 \\ 0 & 0 & x_2 & y_2 \\ 0 & 0 & z_2 & w_2 \end{pmatrix}, \qquad (0,0)(0,1) = \begin{pmatrix} x_1/\sqrt{2} & y_1 & x_1/\sqrt{2} & 0 \\ z_1/\sqrt{2} & w_1 & z_1/\sqrt{2} & 0 \\ -x_2/\sqrt{2} & 0 & x_2/\sqrt{2} & y_2 \\ -z_2/\sqrt{2} & 0 & z_2/\sqrt{2} & w_2 \end{pmatrix},$$

$$(0,1)(0,0) = \begin{pmatrix} x_1/\sqrt{2} & y_1/\sqrt{2} & -x_2/\sqrt{2} & -y_2/\sqrt{2} \\ z_1 & w_1 & 0 & 0 \\ x_1/\sqrt{2} & y_1/\sqrt{2} & x_2/\sqrt{2} & y_2/\sqrt{2} \\ 0 & 0 & z_2 & w_2 \end{pmatrix},$$

$$(1,0)(0,0) = \begin{pmatrix} x_1 & y_1 & 0 & 0 \\ z_1/\sqrt{2} & w_1/\sqrt{2} & -z_2/\sqrt{2} & -w_2/\sqrt{2} \\ 0 & 0 & x_2 & y_2 \\ z_1/\sqrt{2} & w_1/\sqrt{2} & z_2/\sqrt{2} & w_2/\sqrt{2} \end{pmatrix},$$

$$(1,1)(0,0) = \begin{pmatrix} x_1/\sqrt{2} & y_1/\sqrt{2} & -x_2/\sqrt{2} & -y_2/\sqrt{2} \\ z_1/\sqrt{2} & w_1/\sqrt{2} & -z_2/\sqrt{2} & -w_2/\sqrt{2} \\ x_1/\sqrt{2} & y_1/\sqrt{2} & x_2/\sqrt{2} & y_2/\sqrt{2} \\ z_1/\sqrt{2} & w_1/\sqrt{2} & z_2/\sqrt{2} & w_2/\sqrt{2} \end{pmatrix},$$

$$(0,1)(0,1) = \begin{pmatrix} (x_1+x_2)/2 & y_1/\sqrt{2} & (x_1-x_2)/2 & -y_2/\sqrt{2} \\ z_1/\sqrt{2} & y_1/\sqrt{2} & z_1/\sqrt{2} & 0 \\ (x_1-x_2)/2 & -w_1/\sqrt{2} & (x_1+x_2)/2 & y_2/\sqrt{2} \\ -z_2/\sqrt{2} & 0 & z_2/\sqrt{2} & w_2/\sqrt{2} \end{pmatrix},$$

$$(1,0)(0,1) = \begin{pmatrix} x_1/\sqrt{2} & y_1 & x_1/\sqrt{2} & 0 \\ (z_1+z_2)/2 & w_1/\sqrt{2} & (z_1-z_2)/2 & -w_2/\sqrt{2} \\ -x_1/\sqrt{2} & 0 & x_2/\sqrt{2} & y_2 \\ (z_1-z_2)/2 & w_1/\sqrt{2} & (z_1+z_2)/2 & w_2/\sqrt{2} \end{pmatrix},$$

$$(1,1)(0,1) = \begin{pmatrix} (x_1+x_2)/2 & y_1/\sqrt{2} & (x_1-x_2)/2 & -y_2/\sqrt{2} \\ (z_1+z_2)/2 & w_1/\sqrt{2} & (z_1-z_2)/2 & -w_2/\sqrt{2} \\ (x_1-x_2)/2 & y_1/\sqrt{2} & (x_1+x_2)/2 & y_2/\sqrt{2} \\ (z_1-z_2)/2 & w_1/\sqrt{2} & (z_1+z_2)/2 & w_2/\sqrt{2} \end{pmatrix},$$

$$(0,0)(1,0) = \begin{pmatrix} x_1 & y_1/\sqrt{2} & 0 & y_1/\sqrt{2} \\ z_1 & w_1/\sqrt{2} & 0 & w_1/\sqrt{2} \\ 0 & -y_2/\sqrt{2} & x_2 & y_2/\sqrt{2} \\ 0 & -w_2/\sqrt{2} & z_2 & w_2/\sqrt{2} \end{pmatrix}, \quad (0,0)(1,1) = \frac{1}{\sqrt{2}}\begin{pmatrix} x_1 & y_1 & x_1 & y_1 \\ z_1 & w_1 & z_1 & w_1 \\ -x_2 & -y_2 & x_2 & y_2 \\ -z_2 & -w_2 & z_2 & w_2 \end{pmatrix},$$

$$(0,1)(1,0) = \begin{pmatrix} x_1/\sqrt{2} & (y_1+y_2)/2 & -x_2/\sqrt{2} & (y_1-y_2)/2 \\ z_1 & w_1/\sqrt{2} & 0 & w_1/\sqrt{2} \\ x_1/\sqrt{2} & (y_1-y_2)/2 & x_2/\sqrt{2} & (y_1+y_2)/2 \\ 0 & -w_2/\sqrt{2} & z_2 & w_2/\sqrt{2} \end{pmatrix},$$

$$(1,0)(1,0) = \begin{pmatrix} x_1 & y_1/\sqrt{2} & 0 & y_1/\sqrt{2} \\ z_1/\sqrt{2} & (w_1+w_2)/2 & -z_2/\sqrt{2} & (w_1-w_2)/2 \\ 0 & -y_2/\sqrt{2} & x_2 & y_2/\sqrt{2} \\ z_1/\sqrt{2} & (w_1-w_2)/2 & z_2/\sqrt{2} & (w_1+w_2)/2 \end{pmatrix},$$

$$(1,1)(1,0) = \begin{pmatrix} x_1/\sqrt{2} & (y_1+y_2)/2 & -x_2/\sqrt{2} & (y_1-y_2)/2 \\ z_1/\sqrt{2} & (w_1+w_2)/2 & -z_2/\sqrt{2} & (w_1-w_2)/2 \\ x_1/\sqrt{2} & (y_1-y_2)/2 & x_2/\sqrt{2} & (y_1+y_2)/2 \\ z_1/\sqrt{2} & (w_1-w_2)/2 & z_2/\sqrt{2} & (w_1+w_2)/2 \end{pmatrix},$$

$$(0,1)(1,1) = \begin{pmatrix} (x_1+x_2)/2 & (y_1+y_2)/2 & (x_1-x_2)/2 & (y_1-y_2)/2 \\ z_1/\sqrt{2} & w_1/\sqrt{2} & z_1/\sqrt{2} & w_1/\sqrt{2} \\ (x_1-x_2)/2 & (y_1-y_2)/2 & (x_1+x_2)/2 & (y_1+y_2)/2 \\ -z_2/\sqrt{2} & -w_2/\sqrt{2} & z_2/\sqrt{2} & w_2/\sqrt{2} \end{pmatrix},$$

$$(1,0)(1,1) = \begin{pmatrix} x_1/\sqrt{2} & y_1/\sqrt{2} & x_1/\sqrt{2} & y_1/\sqrt{2} \\ (z_1+z_2)/2 & (w_1+w_2)/2 & (z_1-z_2)/2 & (w_1-w_2)/2 \\ -x_2/\sqrt{2} & -y_2/\sqrt{2} & x_2/\sqrt{2} & y_2/\sqrt{2} \\ (z_1-z_2)/2 & (w_1-w_2)/2 & (z_1+z_2)/2 & (w_1+w_2)/2 \end{pmatrix},$$

$$(1,1)(1,1) = \frac{1}{2}\begin{pmatrix} x_1+x_2 & y_1+y_2 & x_1-x_2 & y_1-y_2 \\ z_1+z_2 & w_1+w_2 & z_1-z_2 & w_1-w_2 \\ x_1-x_2 & y_1-y_2 & x_1+x_2 & y_1+y_2 \\ z_1-z_2 & w_1-w_2 & z_1+z_2 & w_1+w_2 \end{pmatrix}.$$